\documentclass[aps,prl,twocolumn,showpacs,superscriptaddress]{revtex4}

\usepackage{amsmath,bm}
\usepackage{graphicx}
\usepackage{upgreek}

\newcommand{\nTrtHz}{\mathrm{nT}/\sqrt{\mathrm{Hz}}}
\newcommand{\sigmaIR}{\sigma_{\mathrm{IR}}}
\newcommand{\nNV}{n_{\mathrm{NV}}}
\newcommand{\NNV}{N_{\mathrm{NV}}}
\newcommand{\SLI}{S_{\mathrm{LI}}}

\newcommand{\lr}[1]{\left(  #1 \right)}
\newcommand{\abs}[1]{\left|  #1 \right|}
\newcommand{\rtHz}{\sqrt{\mathrm{Hz}}}

\newcommand{\Psat}{P_{\mathrm{sat}}}

\newcommand{\fdev}{f_{\mathrm{dev}}}
\newcommand{\fMW}{f_{\mathrm{MW}}}
\newcommand{\fmod}{f_{\mathrm{mod}}}
\newcommand{\fres}{f_{\mathrm{res}}}

\begin{document}
 
\title{Cavity-enhanced room-temperature magnetometry using absorption by nitrogen-vacancy centers in diamond}

\author{K. Jensen}
\affiliation{Department of Physics, University of California, Berkeley, California 94720-7300, USA}
\author{N. Leefer }
\affiliation{Department of Physics, University of California, Berkeley, California 94720-7300, USA}
\author{ A. Jarmola }
\affiliation{Department of Physics, University of California, Berkeley, California 94720-7300, USA}
\author{Y. Dumeige}
\affiliation{UEB, Universit$\acute{e}$ Europ$\acute{e}$enne de Bretagne, Universit$\acute{e}$ de Rennes I}
\affiliation{CNRS, UMR 6082 FOTON, Enssat, 6 rue de Kerampont, CS 80518, 22305 Lannion cedex, France}
\author{V. M. Acosta }
\affiliation{Department of Physics, University of California, Berkeley, California 94720-7300, USA}
\author{P. Kehayias }
\affiliation{Department of Physics, University of California, Berkeley, California 94720-7300, USA}
\author{B. Patton}
\affiliation{Department of Physics, University of California, Berkeley, California 94720-7300, USA}
\affiliation{Physik-Department, Technische Universit\"{a}t M\"{u}nchen, 85748 Garching, Germany}
\author{D. Budker}
\affiliation{Department of Physics, University of California, Berkeley, California 94720-7300, USA}

\begin{abstract}
We demonstrate a cavity-enhanced room-temperature magnetic field sensor based on nitrogen-vacancy centers in diamond. 
Magnetic resonance is detected using absorption of light resonant with the 1042 nm spin-singlet transition. 
The diamond is placed in an external optical cavity to enhance the absorption, and significant absorption is observed even at room temperature. 
We  demonstrate a magnetic field sensitivity of 2.5~$\mathrm{nT}/\sqrt{\mathrm{Hz}}$, and project a photon shot-noise-limited sensitivity of 70~$\mathrm{pT}/\sqrt{\mathrm{Hz}}$ for 
a few mW of infrared light, and a quantum projection-noise-limited sensitivity of
250~$\mathrm{fT}/\sqrt{\mathrm{Hz}}$
for the sensing volume of
$\sim 90 \ \upmu \mathrm{ m} \times 90 \ \upmu \mathrm{ m} \times 200 \ \upmu \mathrm{ m}$.

% 07.55.Ge 	Magnetometers for magnetic field measurements 
% 76.30.Mi 	Color centers and other defects, EPR 
% 81.05.ug 	Diamond

% 61.72.jn 	Color centers, crystal defects
% 76.70.Hb 	Optically detected magnetic resonance (ODMR) 
% Sensors, magnetic field, 85.75.Ss
% 76.60.Es 	Relaxation effects 

\end{abstract}
\maketitle

Optical sensing of magnetic fields \cite{Taylor08}, electric fields \cite{Dolde2011, Acosta2013}, rotations \cite{Maclaurin2012, Ledbetter2012,Ashok2012}, and temperature \cite{Acosta2010,Kuscko2013,Toyli2013,Neumann2013} can be achieved using negatively charged nitrogen-vacancy (NV) centers in diamond. 
Single NV centers and ensembles of NV centers can be detected with high spatial resolution and can be used as sensors with nm, $\upmu$m, or mm resolution \cite{Balasubramanian08, Maze08, Rittweger2009,Acosta09PRB}.
Most of these sensors are based on fluorescence detection of the NV center's spin state and suffer from low photon detection efficiency and background fluorescence. Even with improved photon collection  \cite{Hadden10,Siyushev10,Sage12}, current state-of-the-art magnetic field sensors \cite{Sage12} can only reach sensitivities which are several orders of magnitude worse than the quantum projection noise (PN) limited sensitivity \cite{Budker07,Acosta09PRB} associated with the finite number of sensing spins.
Sensors based on detection of infrared (IR) absorption \cite{Acosta10APL} achieve high photon detection efficiency and can reach a sensitivity closer to the PN-limit compared to sensors based on fluorescence detection when a cavity is used to enhance the detection-contrast \cite{Dumeige13}.
Previously, absorption of IR light has been used for magnetometry at cryogenic temperatures ($\sim70$ K) \cite{Acosta10APL} where the absorption is much stronger than at room temperature.
In this work, a diamond plate is placed in an external cavity which enhances the absorption from the NV centers and the spin-state detection-contrast.
The device is used as a magnetic field sensor operated at room-temperature, but could be used for other applications requiring high optical depth such as electromagnetically induced transparency \cite{Acosta2013} or optical quantum memory \cite{Heshami2013arxiv}.

\begin{figure}
\centering
\includegraphics[width=0.4\textwidth]{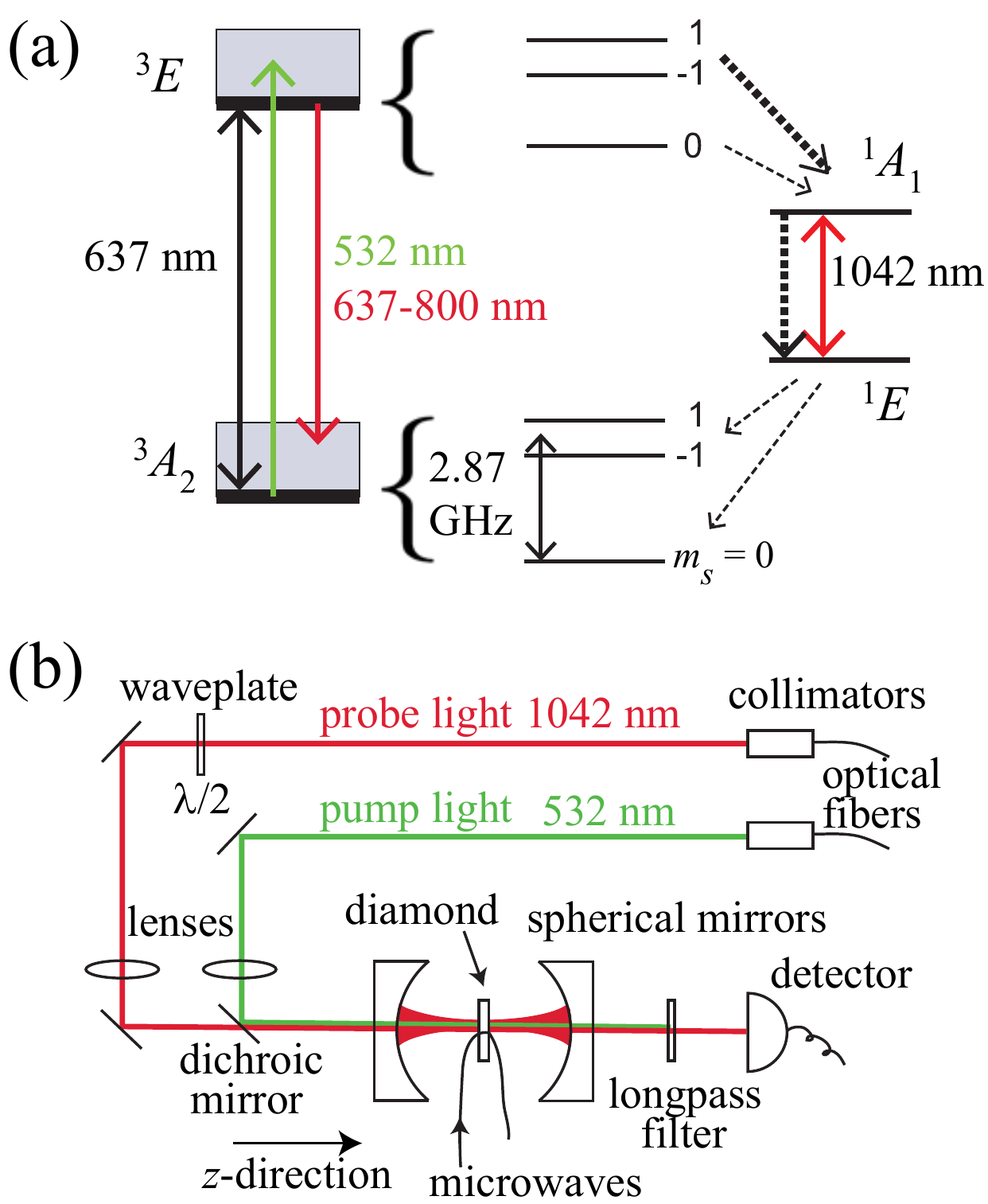}
\caption{(a) Levels and transitions of the NV center. 
Solid lines indicate optical and microwave transitions; dashed lines indicate non-radiative transitions.
(b) Setup. The cavity was placed inside a sound insulation enclosure (not shown).}
\label{fig:levelssetup}
\end{figure}

The level structure of the NV center is shown in Fig.~\ref{fig:levelssetup}(a). 
Electron spin-triplet and spin-singlet states are labeled $^3A_2$, $^3E$ and $^1E$, $^1A_1$, respectively. 
The NV center can be excited optically from the ground state $^3A_2$ to the state $^3E$. 
From the $^3E$ state, the NV center can decay to the $^3A_2$ state through a spin-conserving transition which leads to fluorescence in the 637-800 nm wavelength range.
It can also decay to the upper singlet state $^1A_1$ through a spin-nonconserving transition,
which occurs with higher probability for the $m_s=\pm 1$ states compared to the $m_s=0$ state \cite{Robledo11, Tetienne2012NJP}.
From the $^1A_1$  state, the NV center decays through a 1042 nm transition to the metastable $^1E$ singlet state, which has a lifetime of $\sim 200$ ns at room temperature \cite{Acosta10PRB}.
The NV center then decays from the $^1E$ state back to the $^3A_2$ ground state.
Under continuous illumination with sufficiently strong green pump light at 532 nm, the NV center is mainly in the $^3A_2$  $m_s=0$ ground state and the $^1E$ metastable singlet state.
Due to the spin-dependent transition rates, application of microwaves on resonance with the $m_s=0 \leftrightarrow m_s=\pm 1$ transitions leads to increased population of the $^1E$ metastable singlet state.
Absorption of 1042 nm light can therefore be used as a probe of the transitions within the spin-triplet ground state.

\begin{figure*}
\centering
\includegraphics[width=\textwidth]{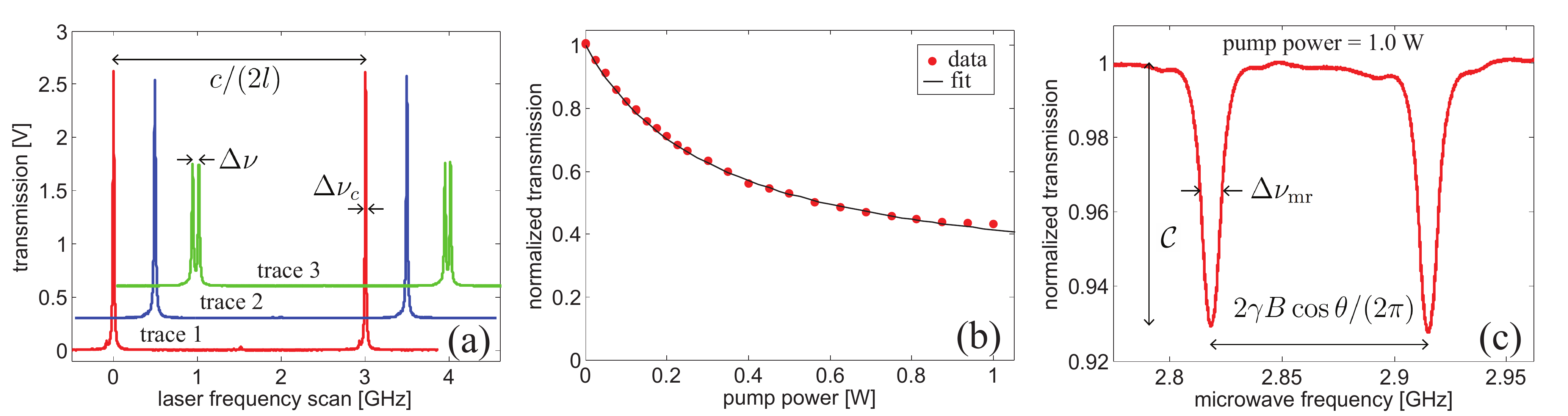}
\caption{
(a) Transmission of IR light through the cavity when the IR laser frequency is scanned. 
Trace 1: the cavity is empty. 
Trace 2 and 3: The diamond is located inside the cavity. 
The experimental settings for trace 2 and 3 are the same except for the input light polarization.
The spectra have been offset and shifted for clarity. The pump light was blocked for these measurements.
(b)~Transmission of IR light (with the laser locked to a cavity resonance) as a function of pump power measured before the cavity. The pump power inside the diamond is lower due to reflections from the front mirror of the cavity and the diamond surface. The transmission is here normalized to unity for zero pump power.
(c)~Transmission 
(normalized to unity off resonance) 
%(normalized to unity when the microwaves are off-resonance) 
as a function of microwave frequency. 
A 2.99~mT magnetic field is aligned along the $z$-direction.
}
\label{fig:transmissioncombined}
\end{figure*}

\begin{figure*}
\centering
\includegraphics[width=\textwidth]{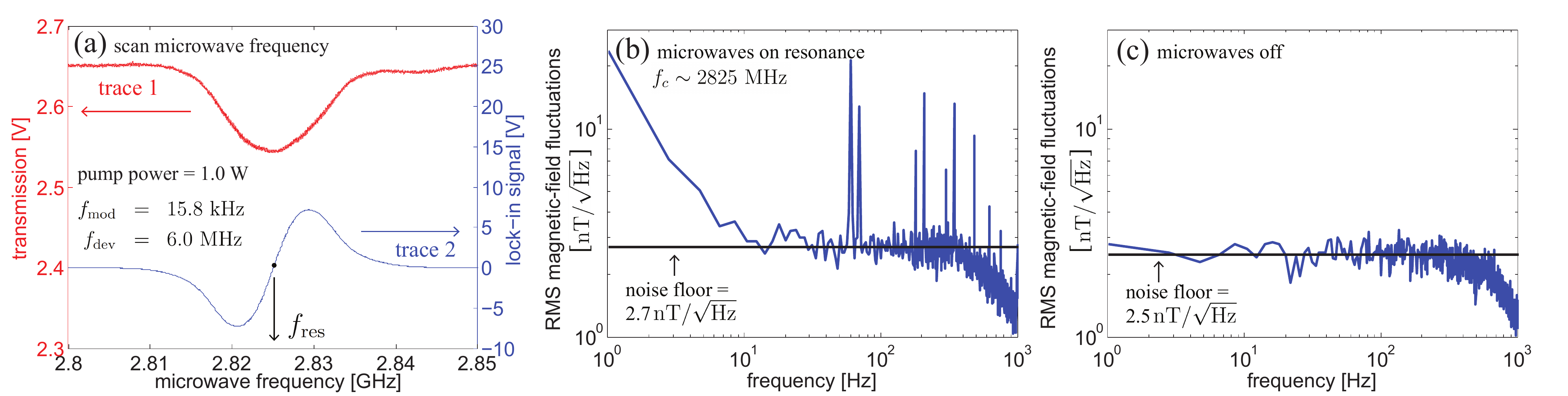}
\caption{
(a) Cavity-transmission signal (top trace) and lock-in signal (bottom trace) as a function of the central microwave frequency $f_c$ which is scanned over a magnetic resonance. 
%The modulation frequency and the frequency deviation were set to $\fmod = 15.8$~kHz and $\fdev = 6.0$~MHz, respectively.
(b)~Magnetic field noise spectrum calculated using the output of the lock-in amplifier with a total measurement time of 10~s. 
The central frequency $f_c$ is tuned to the magnetic resonance $\fres$. 
(c)~Noise spectrum for the case where the microwaves were turned off.
}
\label{fig:noise123}
\end{figure*}

The experimental setup is shown in Fig.~\ref{fig:levelssetup}(b). 
1042 nm light is provided by an external-cavity diode laser and 532 nm light is provided by a diode-pumped solid-state laser.
%Green (532 nm) and IR (1042 nm) laser light is used in the experiment.
The optical cavity consists of two spherical mirrors each with a radius of curvature of $r=5$ cm, an intensity reflectivity of $R=98(1)\%$  for 1042 nm light, and a transmitivity of $T=70(1)\%$ for 532 nm light. 
For the first measurements (shown in Fig.~\ref{fig:transmissioncombined}), we choose the mirror spacing  $l =5.0$~cm resulting in beam waist $w_0=91$~$\upmu$m and Rayleigh range $z_0=2.5$~cm for the 1042 nm $\mathrm{TEM}_{00}$ cavity mode \cite{Lasers}.
For the next measurements (shown in Fig.~\ref{fig:noise123}), we choose $l=2.5$~cm resulting in  $w_0=85$~$\upmu$m and $z_0=2.2$~cm.
The IR laser beam was matched to the $\mathrm{TEM}_{00}$ cavity mode in both cases. The pump laser beam was overlapped with the IR laser beam in the center of the cavity, but it was not mode-matched to the cavity. The waist of the pump beam was $90(2)$~$\upmu$m as measured with a CCD camera.
Trace 1 in Fig.~\ref{fig:transmissioncombined}(a) shows the transmission of IR light through the cavity while scanning the laser frequency. The frequency interval between the two resonances is $ c/\lr{2l} = 3.0$~GHz, where $l= 5.0$~cm and $c$ is the speed of light. From Lorentzian fits to the resonances in trace~1, we find the full width at half maximum (FWHM) of the resonances  $\Delta \nu_c = 14.9(15)$~MHz and the cavity finesse $\mathcal{F} = \left[c/\lr{2l}\right]/\Delta \nu_c=202(20)$.
The measured finesse corresponds to mirror reflectivity $R=98.5(2)\%$ as calculated from Eq.~(S3) in the Supplementary Material \cite{IRsup}.

Diamond has a refractive index of $n=2.4$ which leads to a reflection of $17\%$ for normal incidence at a diamond/air interface. 
Our diamond sample with size 3~mm~$\times$~3~mm~$\times$~0.2~mm is anti-reflection coated and has a measured total reflection from both sides of $0.1\%$ at 1042~nm. The sample was irradiated with relativistic electrons and annealed as in Ref.~\cite{Acosta09PRB} in order to increase the density of NV centers.
Trace 2 in Fig.~\ref{fig:transmissioncombined}(a) shows the transmission through the cavity when the diamond is inserted inside the cavity. The peak transmission decreases slightly compared to an empty cavity and the finesse is lowered to 
$\mathcal{F}=165(17)$ due to additional losses introduced by the diamond. 
This value corresponds to a single-pass loss in the diamond of $0.35\%$, as calculated from Eq.~(S3) with $R=98.5\%$.
The cavity resonance frequencies depend on the index of refraction of the diamond. We observe significant linear birefringence of the diamond \cite{IRsup}, as seen in Trace 3 in Fig.~\ref{fig:transmissioncombined}(a) where twice the number of resonances are present in the spectrum. The large linear birefringence constrains the polarization of the light transmitted through the cavity to be either along or perpendicular to the diamond's axis of birefringence. In the measurements discussed below, we adjusted the input light polarization to maximize one cavity resonance [as in trace 2 in Fig.~\ref{fig:transmissioncombined}(a)] and locked the laser frequency to this resonance.

Absorption of light by an NV center depends on the light polarization, and for both the $^3A_2 \leftrightarrow {^3}E$ triplet transition and the $^1E \leftrightarrow {^1}A_1$ singlet transition, the absorption is maximal when the polarization is perpendicular to the NV axis and zero when the polarization is along the NV axis \cite{Acosta10PRB}.  
The NV center's axis 
\footnote{The NV center's axis corresponds to the vector connecting the nitrogen atom and the neighbouring vacancy.}
can be aligned in four different ways in the diamond crystal, corresponding to the four [111] crystallographic directions. 
Our diamond is cut along one of the (100) planes and positioned such that its [100] axis coincide with the light propagation direction  [$z$-direction in Fig.~\ref{fig:levelssetup}(b)].
In this case, the \emph{total} absorption by all NV centers is relatively insensitive to the polarization of the pump and probe light.
This is an advantage when the diamond is birefringent and the probe polarization cannot be chosen freely. 

Figure~\ref{fig:transmissioncombined}(b) shows the cavity transmission of 1042 nm light as a function of 532 nm pump light power.
The drop in transmission at higher pump powers is attributed to absorption from a larger number of NV centers in the singlet states.
We observe a large ($>50\%$) change in transmission for high pump powers. Such a large change  is possible at room temperature 
because the absorption is enhanced by the factor $2\mathcal{F}/\pi$ due to the cavity compared to a single-pass scheme \cite{IRsup}.
Absorption of IR light due to NV centers can be modeled as a pump power-dependent loss inside the cavity. 
The single-pass transmission through the diamond is $L_0-A_0 P/ \lr{P+\Psat}$, where $L_0$ is the transmission coefficient in the absence of pump light, $P$ is the pump power, $\Psat$ is a saturation pump power,
and $P/\lr{P+\Psat}$ is the fraction of NV centers which are pumped into the metastable singlet state.
The data in Fig.~\ref{fig:transmissioncombined}(b) are fitted to such a model 
[using Eq.~(S2)]
and we find $A_0=2.2(1)\%$ and $\Psat=0.88(3)~$W.
The parameter $A_0$ is related to the density of NV centers through $\nNV \approx A_0/\lr{\sigmaIR\cdot d}$ \cite{IRsup}, where $\sigmaIR=3_{-1}^{+3}\cdot10^{-22}~\rm{m}^2$ 
is the cross-section for IR light \cite{Dumeige13,Kehayias2013} and $d$ is the thickness of the diamond.
We calculate
$\nNV =3.6(1.8)\cdot 10^{23}~\rm{m}^{-3}$ or equivalently 2(1)~ppm which is 
in agreement with the expected value for this type of electron-irradiated diamond \cite{Acosta09PRB}.
% Fitted parameters: {A_0 -> 0.0215307, P0 -> 3.51491} but should use Psat=P_{set Verdi}*0.25
%
The saturation pump power $\Psat$ corresponds to a peak intensity 
$I_0=2\Psat/\lr{\pi w_0^2} =70$ MW/$\mathrm{m}^2$,
which is an order of magnitude smaller than the expected saturation intensity as calculated from the pump light cross-section and the  lifetime of the metastable singlet state \cite{IRsup}. The discrepancy could be due to other pump-power dependent effects such as photo-ionization  of NV centers.

The NV center's ground-state energy levels shift due to the Zeeman effect  in an applied magnetic field.
The Zeeman effect can be described by the Hamiltonian 
$\mathcal{H}_B = \gamma \mathbf{B} \cdot \mathbf{S} $, where $ \gamma = 2\pi \cdot 28.0$~GHz/T is the gyromagnetic ratio for the NV center. A permanent ring magnet is used to apply a magnetic field, $\mathbf{B}$, along the $z$-direction which coincide with a [100] direction. In this case, the angle between any NV center's axis and the magnetic field is $\theta=54.7^\circ$.
For magnetic fields $B \ll 2\pi D/\lr{\gamma \cos \theta}\sim 0.2$~T, where $D=2.87$~GHz is the zero-field splitting  of the NV center's ground state energy levels, the Zeeman shift depends to first order only on the projection of the magnetic field on the NV axis. The shift in Hz is $\approx m_s  \gamma B \cos \theta/\lr{2\pi}$ and is the same for all NV centers, independent of their orientations. Microwaves are applied with a wire positioned on top of the diamond.
Figure~\ref{fig:transmissioncombined}(c) shows the cavity transmission while the microwave frequency is scanned. The pump power was 1.0~W for this measurement. Two magnetic resonances are observed corresponding to the 
$m_s = 0 \leftrightarrow m_s = - 1$ and $m_s = 0 \leftrightarrow m_s = + 1$ transitions within the ground state.
The resonance frequencies are  $\fres=D \pm \gamma B \cos \theta/\lr{2\pi}$ and the magnetic field is determined to be 2.99~mT from the frequency difference between the resonances.
The contrast $\mathcal{C}$ and the FWHM $\Delta \nu_{\mathrm{mr}}$ of the magnetic resonances 
[see Fig.~\ref{fig:transmissioncombined}(c)] depend on the pump and microwave powers \cite{Jensen2013}, which were chosen to optimize the magnetic field sensitivity.
For the resonances in Fig.~\ref{fig:transmissioncombined}(c) we find  $\mathcal{C}=7.1\%$ and $\Delta \nu_{\mathrm{mr}}=9.0$~MHz.
When detecting IR absorption at room temperature, such a high contrast  is only possible due to the cavity enhancement which increases  $\mathcal{C}$ by the factor $2\mathcal{F}/\pi \sim 75$.
The sensitivity to magnetic fields aligned along the $z$-direction is in the absence of technical noise limited by the shot noise (SN) of the probe light. For the spectrum presented in Fig.~\ref{fig:transmissioncombined}(c), we find a SN-limited sensitivity
$
\delta B_{\mathrm{SN}} \approx 
\left( 
2\pi  \Delta \nu_{\rm{mr}}
\right)/
\left(  	
\gamma \mathcal{C}\sqrt{\mathcal{R}} \cos \theta
\right)
$ 
 \cite{Acosta2013}
of 70~pT/${\rtHz}$ using a detected IR power of 2.3~mW ($\mathcal{R}$ is the rate of detected photons).

In order to make a sensitive measurement of the magnetic field, we utilize a technique where the microwave frequency $\fMW$ is modulated around a central frequency $f_c$:
$f_{\mathrm{MW}} = f_c + \fdev \cos\lr{2 \pi \fmod t}$, where
 $\fmod$ is the modulation frequency and $\fdev$ is the amplitude of modulation.
The detected signal (the transmission through the cavity) is demodulated with a lock-in amplifier referenced to the modulation frequency $\fmod$. The output of the lock-in amplifier is linear $ \SLI \approx  \alpha \lr{f_c-\fres}$ as a function of $f_c-\fres$ 
when  $\abs{f_c -\fres}  < \Delta \nu_{\mathrm{mr}}/2 $.
Besides providing a linear magnetometer signal, the modulation technique reduces the sensitivity to technical low-frequency noise.
Trace 1 in Fig.~\ref{fig:noise123}(a) shows the transmission while $f_c$ is scanned over the 
lowest-frequency resonance in Fig.~\ref{fig:transmissioncombined}(c). The fast modulation is not visible in this trace due to averaging. 
The slope $\alpha\propto \mathcal{C}/\Delta\nu_{\mathrm{mr}}$ is extracted from trace~2 in Fig.~\ref{fig:noise123}(a) which shows the output of the lock-in amplifier. Due to the finite bandwidth of the magnetometer (13.5~kHz, see Fig.~S1), the slope $\alpha$ also depends on $\fmod$. 
For $\fmod=15.8$~kHz, $\alpha$ was reduced by a factor $\sim 2.2$ compared to low modulation frequencies. 
We chose  $\fmod=15.8$~kHz since this maximized the measured signal-to-noise ratio. 
For $\fmod<10$~kHz, intensity noise of the pump laser became a dominant noise source.

The lock-in signal is linear in small changes $\Delta B$ in magnetic field around $B_0$, when the central microwave frequency is set to $f_c=D-\gamma B_0 \cos \theta /\lr{2\pi}$. Under this condition, the output of the lock-in amplifier was recorded  for 10~s. 
Knowing the slope $\alpha$, the change in magnetic field 
$\Delta B(t) = -2\pi  \cdot \SLI(t) /  \lr{\alpha \gamma\cos \theta }$
as a function of time can be calculated.
The root-mean-square (RMS) magnetic field fluctuations 
[which equals the square-root of the power spectral density of $\Delta B(t)$]
are shown in Fig.~\ref{fig:noise123}(b).
For frequencies in the range 10-500~Hz we reach a noise floor of 2.7~$\nTrtHz$. 
For frequencies below 10 Hz additional noise is present.
Above 500~Hz the noise decreases due to the finite time constant of the lock-in amplifier (here 320~$\upmu$s), which reduces the bandwidth to $\sim1$~kHz. Several noise peaks are also present.

The magnetic-resonance position depends on the temperature $\mathcal{T}$, since around room temperature the zero-field splitting $D$ depends linearly on $\mathcal{T}$ with the slope  $dD/d\mathcal{T} =-74$~kHz/K  \cite{Acosta2010}.
The sensitivity of our magnetometer to temperature changes is 
$ \gamma \cos \theta/\lr{2\pi \cdot dD/d\mathcal{T}} = 0.22$~K/$\upmu$T.
Fig.~\ref{fig:noise123}(c)  show the noise spectrum when the microwaves are off, where the absorption due to NV centers does not depend on  magnetic field or temperature.
In this case we do not observe the additional low-frequency noise or the noise peaks. 
Based on this observation and additional measurements performed with a commercial magnetometer (Honeywell HMC2003), we determine that the noise peaks are due to magnetic noise in the lab while the low-frequency noise is due to temperature fluctations of the diamond.
The high pump power leads to significant heating of the diamond (around 40~K increase for 1.0~W pump power as measured from the shift in resonance frequency).
Thermal fluctuations could arise if the diamond is not in thermal equilibrium or if the pump power fluctuates.
The low-frequency noise could be reduced by improving the thermal contact of the diamond to the mount, by actively stabilizing the temperature of the diamond, or by reducing or stabilizing the pump power.
Alternatively, temperature changes  could be distinguished from magnetic field changes by monitoring the positions of both resonances in Fig.~\ref{fig:transmissioncombined}(c) or by using a quantum beat technique \cite{Kejie2013}.

The noise floor in Fig.~\ref{fig:noise123}(c) is 2.5~$\nTrtHz$ and is limited by laser frequency noise and/or vibrations of the cavity. 
The cavity length was 2.5~cm for the measurements presented in Fig.~\ref{fig:noise123} and 5~cm for the measurements presented in Fig.~\ref{fig:transmissioncombined}.
The  cavity resonance width [$\Delta \nu_c$ in Fig.~\ref{fig:transmissioncombined}(a)] is larger for the shorter cavity.
A short cavity is preferable, since in this case the magnetometer will be less sensitive to laser frequency noise.
We observed an improvement of the noise floor with the shorter cavity, demonstrating that laser frequency noise was limiting the sensitivity for the longer cavity.
We believe that laser frequency noise still limits the sensitivity for the shorter cavity and that further reduction of the cavity length would be beneficial. The contribution from the probe laser intensity noise to the sensitivity is $\sim 0.8~\nTrtHz$ for $\fmod=15.8$~kHz.

We have demonstrated a cavity-enhanced, room-temperature magnetometer based on IR absorption using NV centers in diamond. 
We reach a noise floor of $2.5 \ \mathrm{nT}/\rtHz$ and  project a SN-limited sensitivity of $70~\mathrm{pT}/\rtHz$ for our current apparatus. The SN-limited sensitivity could be improved by increasing the probe power. 
The 1042 nm transition saturation intensity was predicted to be $\sim 500$~GW/$\mathrm{m}^2$ \cite{Dumeige13}, which for a $90$~$\upmu$m beam waist corresponds to a saturation power of 6.4~kW. Therefore, saturation effects will not be important. 
A realistic increase of the probe power by a factor of 100 would increase the SN-limited sensitivity by a factor of 10 to $\sim 7~\mathrm{pT}/\rtHz$.
Detection of magnetic resonance using IR absorption enhanced by a cavity has the advantage that 100\% contrast is possible using a critically coupled cavity \cite{Dumeige13},  and this method is in principle more sensitive than methods based on detecting fluorescence \cite{Dumeige13}. 
For any detection method, the sensitivity is fundamentally limited by the quantum projection noise  \cite{Budker07,Acosta09PRB} 
$\delta B_{\mathrm{PN}} \approx \lr{2\pi}/\lr{\gamma \sqrt{\NNV T_2}} \sim 250$~fT/$\rtHz$, 
where $\NNV$ is the number of NV centers  and $T_2=1/\lr{\pi \Delta \nu_{\mathrm{mr}}}$ is the coherence time.
A sensitivity in this range is feasible using a critically coupled cavity and an IR power of a few hundreds of mW.
Optimizing the diamond sample in terms of density of NV centers, substitutional nitrogen atoms, and $^{13}$C nuclei could together with dynamical decoupling sequences \cite{Taylor08,BarGill2012} lead to narrower magnetic resonances and even higher sensitivity.

This research was supported by the AFOSR/DARPA QuASAR program and by NSF through Grant No. ECCS-1202258.
K. J. was supported by the Danish Council for Independent Research $|$ Natural Sciences and the Carlsberg Foundation.
Y. D. acknowledges support from Institut Universitaire de France.
We thank G. Vasilakis for commenting on the manuscript.

%\bibliography{BIBInfraredAbsorption} 

\begin{thebibliography}{10}

\bibitem{Taylor08}
J.~M. Taylor, P.~Cappellaro, L.~Childress, L.~Jiang, D.~Budker, P.~R. Hemmer,
  A.~Yacoby, R.~Walsworth, and M.~D. Lukin.
%\newblock High-sensitivity diamond magnetometer with nanoscale resolution.
\newblock {\em Nature Physics}, 4:810, 2008.

\bibitem{Dolde2011}
F.~Dolde, H.~Fedder, M.~W. Doherty, T.~N\"{o}bauer, F.~Rempp,
  G.~Balasubramanian, T.~Wolf, F.~Reinhard, L.~C.~L. Hollenberg, F.~Jelezko,
  and J.~Wrachtrup.
%\newblock Electric-field sensing using single diamond spins.
\newblock {\em Nature Physics}, 7:459, 2011.

\bibitem{Acosta2013}
V.~M. Acosta, K.~Jensen, C.~Santori, D.~Budker, and R.~G. Beausoleil.
%\newblock Electromagnetically induced transparency in a diamond spin ensemble enables all-optical electromagnetic field sensing.
\newblock {\em Phys. Rev. Lett.}, 110:213605, 2013.

\bibitem{Maclaurin2012}
D.~Maclaurin, M.~W. Doherty, L.~C.~L. Hollenberg, and A.~M. Martin.
%\newblock Measurable quantum geometric phase from a rotating single spin.
\newblock {\em Phys. Rev. Lett.}, 108:240403, 2012.

\bibitem{Ledbetter2012}
M.~P. Ledbetter, K.~Jensen, R.~Fischer, A.~Jarmola, and D.~Budker.
%\newblock Gyroscopes based on nitrogen-vacancy centers in diamond.
\newblock {\em Phys. Rev. A}, 86:052116, 2012.

\bibitem{Ashok2012}
Ashok Ajoy and Paola Cappellaro.
%\newblock Stable three-axis nuclear-spin gyroscope in diamond.
\newblock {\em Phys. Rev. A}, 86:062104, 2012.

\bibitem{Acosta2010}
V.~M. Acosta, E.~Bauch, M.~P. Ledbetter, A.~Waxman, L.-S. Bouchard, and
  D.~Budker.
%\newblock Temperature dependence of the nitrogen-vacancy magnetic resonance in diamond.
\newblock {\em Phys. Rev. Lett.}, 104:070801, 2010.

\bibitem{Kuscko2013}
G.~Kucsko, P.~C. Maurer, N.~Y. Yao, M.~Kubo, H.~J. Noh, P.~K. Lo, H.~Park, and
  M.~D. Lukin.
%\newblock Nanometre-scale thermometry in a living cell.
\newblock {\em Nature}, 500:54, 2013.

\bibitem{Toyli2013}
D.~M. Toyli, C.~F. de~las Casas, D.~J. Christle, V.~V. Dobrovitski, and D.~D.
  Awschalom.
%\newblock Fluorescence thermometry enhanced by the quantum coherence of single spins in diamond.
\newblock {\em Proceedings of the National Academy of Sciences},
  110(21):8417, 2013.

\bibitem{Neumann2013}
P.~Neumann, I.~Jakobi, F.~Dolde, C.~Burk, R.~Reuter, G.~Waldherr, J.~Honert,
  T.~Wolf, A.~Brunner, J.~H. Shim, D.~Suter, H.~Sumiya, J.~Isoya, and
  J.~Wrachtrup.
%\newblock High-precision nanoscale temperature sensing using single defects in diamond.
\newblock {\em Nano Letters}, 13:2738, 2013.

\bibitem{Balasubramanian08}
G.~Balasubramanian, I.~Y. Chan, R.~Kolesov, M.~Al-Hmoud, J.~Tisler, C.~Shin,
  C.~Kim, A.~Wojcik, P.~R. Hemmer, A.~Krueger, T.~Hanke, A.~Leitenstorfer,
  R.~Bratschitsch, F.~Jelezko, and J.~Wrachtrup.
%\newblock Nanoscale imaging magnetometry with diamond spins under ambient conditions.
\newblock {\em Nature}, 455:648, 2008.

\bibitem{Maze08}
J.~R. Maze, P.~L. Stanwix, J.~S. Hodges, S.~Hong, J.~M. Taylor, P.~Cappellaro,
  L.~Jiang, M.~V.~Gurudev Dutt, E.~Togan, A.~S. Zibrov, A.~Yacoby, R.~L.
  Walsworth, and M.~D. Lukin.
%\newblock Nanoscale magnetic sensing with an individual electronic spin in diamond.
\newblock {\em Nature}, 455:644, 2008.

\bibitem{Rittweger2009}
Eva Rittweger, Kyu~Young Han, Scott~E. Irvine, Christian Eggeling, and
  Stefan~W. Hell.
%\newblock {STED} microscopy reveals crystal colour centres with nanometric resolution.
\newblock {\em Nature Photonics}, 3:144, 2009.

\bibitem{Acosta09PRB}
V.~M. Acosta, E.~Bauch, M.~P. Ledbetter, C.~Santori, K.-M.~C. Fu, P.~E.
  Barclay, R.~G. Beausoleil, H.~Linget, J.~F. Roch, F.~Treussart,
  S.~Chemerisov, W.~Gawlik, and D.~Budker.
%\newblock Diamonds with a high density of nitrogen-vacancy centers for magnetometry applications.
\newblock {\em Phys. Rev. B}, 80:115202, 2009.

\bibitem{Hadden10}
J.~P. Hadden, J.~P. Harrison, A.~C. Stanley-Clarke, L.~Marseglia, Y.-L.~D. Ho,
  B.~R. Patton, J.~L. O'Brien, and J.~G. Rarity.
%\newblock Strongly enhanced photon collection from diamond defect centers under microfabricated integrated solid immersion lenses.
\newblock {\em Applied Physics Letters}, 97:241901, 2010.

\bibitem{Siyushev10}
P.~Siyushev, F.~Kaiser, V.~Jacques, I.~Gerhardt, S.~Bischof, H.~Fedder,
  J.~Dodson, M.~Markham, D.~Twitchen, F.~Jelezko, and J.~Wrachtrup.
%\newblock Monolithic diamond optics for single photon detection.
\newblock {\em Applied Physics Letters}, 97:241902, 2010.

\bibitem{Sage12}
D.~Le~Sage, L.~M. Pham, N.~Bar-Gill, C.~Belthangady, M.~D. Lukin, A.~Yacoby,
  and R.~L. Walsworth.
%\newblock Efficient photon detection from color centers in a diamond optical waveguide.
\newblock {\em Phys. Rev. B}, 85:121202, 2012.

\bibitem{Budker07}
Dmitry Budker and Michael Romalis.
%\newblock Optical magnetometry.
\newblock {\em Nature Physics}, 3:227, 2007.

\bibitem{Acosta10APL}
V.~M. Acosta, E.~Bauch, A.~Jarmola, L.~J. Zipp, M.~P. Ledbetter, and D.~Budker.
%\newblock Broadband magnetometry by infrared-absorption detection of nitrogen-vacancy ensembles in diamond.
\newblock {\em Appl. Phys. Lett.}, 97:174104, 2010.

\bibitem{Dumeige13}
Y.~Dumeige, M.~Chipaux, V.~Jacques, F.~Treussart, J.-F. Roch, T.~Debuisschert,
  V.~M. Acosta, A.~Jarmola, K.~Jensen, P.~Kehayias, and D.~Budker.
%\newblock Magnetometry with nitrogen-vacancy ensembles in diamond based on infrared absorption in a doubly resonant optical cavity.
\newblock {\em Phys. Rev. B}, 87:155202, 2013.

\bibitem{Heshami2013arxiv}
K.~Heshami, C.~Santori, B.~Khanaliloo, C.~Healey, V.~M. Acosta, P.~E. Barclay,
  and C.~Simon.
%\newblock Raman quantum memory based on an ensemble of nitrogen-vacancy centers coupled to a microcavity.
\newblock arXiv:1312.5342v1, 2013.

\bibitem{Robledo11}
L.~Robledo, H.~Bernien, T.~van~der Sar, and R.~Hanson.
%\newblock Spin dynamics in the optical cycle of single nitrogen-vacancy centres   in diamond.
\newblock {\em New Journal of Physics}, 13:025013, 2011.

\bibitem{Tetienne2012NJP}
J-P Tetienne, L~Rondin, P~Spinicelli, M~Chipaux, T~Debuisschert, J-F Roch, and
  V~Jacques.
%\newblock Magnetic-field-dependent photodynamics of single nv defects in   diamond: an application to qualitative all-optical magnetic imaging.
\newblock {\em New Journal of Physics}, 14:103033, 2012.

\bibitem{Acosta10PRB}
V.~M. Acosta, A.~Jarmola, E.~Bauch, and D.~Budker.
%\newblock Optical properties of the nitrogen-vacancy singlet levels in diamond.
\newblock {\em Phys. Rev. B}, 82:201202, 2010.

\bibitem{Lasers}
Peter~W. Milonni and Joseph~H. Eberly.
\newblock {\em Lasers}.
\newblock Wiley, 1988.

\bibitem{IRsup}
See Supplementary Material.

\bibitem{Kehayias2013}
P.~Kehayias, M.~W. Doherty, D.~English, R.~Fischer, A.~Jarmola, K.~Jensen,
  N.~Leefer, P.~Hemmer, N.~B. Manson, and D.~Budker.
%\newblock Infrared absorption band and vibronic structure of the   nitrogen-vacancy center in diamond.
\newblock {\em Phys. Rev. B}, 88:165202, 2013.

\bibitem{Jensen2013}
K.~Jensen, V.~M. Acosta, A.~Jarmola, and D.~Budker.
%\newblock Light narrowing of magnetic resonances in ensembles of   nitrogen-vacancy centers in diamond.
\newblock {\em Phys. Rev. B}, 87:014115, 2013.

\bibitem{Kejie2013}
K.~Fang, V.~M. Acosta, C.~Santori, Z.~Huang, K.~M. Itoh, H.~Watanabe,
  S.~Shikata, and R.~G. Beausoleil.
%\newblock High-sensitivity magnetometry based on quantum beats in diamond   nitrogen-vacancy centers.
\newblock {\em Phys. Rev. Lett.}, 110:130802, 2013.

\bibitem{BarGill2012}
N.~Bar-Gill, L.M. Pham, A.~Jarmola, D.~Budker, and R.L. Walsworth.
%\newblock Solid-state electronic spin coherence time approaching one second.
\newblock {\em Nature Communications}, 4:1743, 2012.

\end{thebibliography}
%\bibliographystyle{unsrt} 

%%%%%%%%%%%%%% bibliography %%%%%%%%%%%%%%%

%%%%%%%%%%%%%%%%%%%%%%%%%%%%%%%%
\end{document}

\title{Supplementary material to ``Cavity-enhanced room-temperature magnetometry using absorption by nitrogen-vacancy centers in diamond"}

\author{K. Jensen}
\affiliation{Department of Physics, University of California, Berkeley, California 94720-7300, USA}
\author{N. Leefer }
\affiliation{Department of Physics, University of California, Berkeley, California 94720-7300, USA}
\author{ A. Jarmola }
\affiliation{Department of Physics, University of California, Berkeley, California 94720-7300, USA}
\author{Y. Dumeige}
\affiliation{UEB, Universit$\acute{e}$ Europ$\acute{e}$enne de Bretagne, Universit$\acute{e}$ de Rennes I}
\affiliation{CNRS, UMR 6082 FOTON, Enssat, 6 rue de Kerampont, CS 80518, 22305 Lannion cedex, France}
\author{V. M. Acosta }
\affiliation{Department of Physics, University of California, Berkeley, California 94720-7300, USA}
\author{P. Kehayias }
\affiliation{Department of Physics, University of California, Berkeley, California 94720-7300, USA}
\author{B. Patton}
\affiliation{Department of Physics, University of California, Berkeley, California 94720-7300, USA}
\affiliation{Physik-Department, Technische Universit\"{a}t M\"{u}nchen, 85748 Garching, Germany}
\author{D. Budker}
\affiliation{Department of Physics, University of California, Berkeley, California 94720-7300, USA}

\maketitle

\section{Cavity transmission}
Consider an optical cavity consisting of two identical mirrors with intensity transmission and reflection coefficients $T$ and $R$ such that $T+R=1$. An absorptive element  with a single-pass intensity transmission coefficent $L$ is placed inside the cavity. The intensity transmission through the cavity equals \cite{Lasers}
\begin{equation}
I/I_0 = \frac{T^2 L}{(1 - R L)^2 + 4 R L  \lr{\sin \phi}^2 },
\end{equation}
where $I_0$ is the input intensity and $\phi$ is the single-pass phase aquired by the light inside the cavity.
In our experiment the mirrors are curved, and the input light is matched to the $\rm{TEM}_{00}$ cavity mode.
When scannning the frequency of the light, maxima in the transmission will be spaced by the frequency interval $c/\lr{2 \lopt}$, where $\lopt$ is the optical path length of the cavity and  $c$ the speed of light. 
The maximum transmission equals 
\begin{equation}
\Imax/I_0 = \frac{T^2 L}{(1 - R L)^2 }.
\label{eq:Imax}
\end{equation}
The finesse $\mathcal{F}$ of the cavity is defined as the frequency spacing $c/\lr{2 \lopt}$ divided by the cavity resonance FWHM $\Delta\nu_c$  and is given by:
\begin{equation}
\mathcal{F}  = \left[c/\lr{2 \lopt} \right]/\Delta\nu_c \approx  \pi \frac{\sqrt{R L}}{1-R L}.
\label{eq:finesse}
\end{equation}
We are interested in measuring the loss inside the cavity.
Suppose there is a small change $|\delta L|=L_0-L$ in the parameter $L$ from its initial value $L_0$.
By Taylor expanding Eq.~(\ref{eq:Imax}) we find the change in the cavity transmission 
$\Delta I = \Imax \lr{L=L_0} - \Imax \lr{L=L0-|\delta L|}$
to be 
\begin{equation}
\frac{\Delta I}{\Imax\lr{L=L_0}} = 
 \frac{\lr{1+RL_0}}{\lr{1-RL_0}}\frac{|\delta L|}{L_0} 
\approx \frac{2\mathcal{F}}{\pi} \frac{|\delta L|}{L_0}  .
\end{equation}
We see that the sensitivity to a small change in absorption is increased by the factor $2\mathcal{F}/\pi$ by employing a cavity compared to a single pass where $\Delta I/I(L=L_0)\approx |\delta L|/L_0$.
The effective number of passes through the absorptive element inside the cavity is therefore $2\mathcal{F}/\pi$.

\section{Birefringence}
Assume the cavity contains a diamond with thickness $d$ and index of refraction $n=2.4$ such that the total length of the cavity is  $l+d$  and the  optical path length of the cavity is $\lopt=l+nd$.
Light input to the cavity will be resonantly transmitted if the frequency of the light equals a cavity resonance frequency:
\begin{equation}
\nu_m = m \cdot \frac{c}{2 \lr{l+nd}}+\nu_G,%\ \mathrm{where} \ m=0,1,2,... 
\end{equation}
where $m=0,1,2,...$ and $\nu_G$ is related to the Gouy phase shift of a Gaussian beam \cite{Lasers}.
We now assume that the diamond is birefringent such that two orthogonal linear polarizations, here denoted $H$ and $V$, have slightly different index of refraction: $n_H=n$ and $n_V=n+\Delta n$.
The resonance frequencies for H and V polarizations differ by the amount
\begin{equation}
\nu_m^H-\nu_k^V= \frac{m-k}{k} \cdot \nu_k^V + \frac{\Delta n \cdot  d }{l+\lr{n+\Delta n}d} \cdot  \nu_m^H.
\end{equation}
The frequency difference $\nu_m^H-\nu_k^V$ for $H$ and $V$ polarized light is measured in the experiment 
[see trace 3 in Fig.~2(a)]. We believe that neighboring resonances for $H$ and $V$ polarized light have the same fringe order $m=k$. This is based on the observation that by moving the diamond (and thereby probing different locations on the diamond which have different degrees of birefringence), neighboring $H$ and $V$ resonances can be made to overlap but not cross. Assuming $m=k$ and $\Delta n \ll n$ we find  (dropping the subscripts)
\begin{equation}
\Delta \nu \equiv \nu^H-\nu^V \approx \frac{\Delta n \cdot  d }{l+nd} \cdot  \nu_0,
\label{eq:birefringence}
\end{equation}
where $\nu_0$ is the laser frequency.
This equation can be used to estimate the degree of birefringence $\Delta n$ of the diamond.\\

We observe significant birefringence in all our diamond samples and find that the degree of birefringence depends on the spot on the diamond. The birefringence is either an intrinsic property of the diamond \cite{Friel2009} or due to stress induced by the mount.
Traces 2 and 3 in Fig.~2(a) show the cavity transmission for two different polarizations of the light input to the cavity. When the input polarization is aligned either along or perpendicular to the diamond's axis of birefringence (trace 2), peaks separated by the $c/\lr{2 \lopt}=3.0$~GHz are observed. If the polarization is 
neither parallel nor perpendicular
to the birefringence axis (trace 3), twice the number of resonances is observed since light polarized parallel and perpendicular to the material's birefringence axis experiences different index of refraction. 
The degree of birefringence, $\Delta n$, can be estimated from the frequency difference of two neighbouring resonances $\Delta \nu$ using Eq.~(\ref{eq:birefringence}).
From trace 3 we find $\Delta \nu = 70~\mathrm{MHz}$,
which leads to a difference in index of refraction $\abs{\Delta n} = 6.1 \cdot 10^{-5}$, where we used $\nu_0 = c/\lambda = 2.88 \cdot 10^{14}~\mathrm{Hz}$ and $d=0.2$~mm. 
This degree of birefringence corresponds to a phase-shift difference between the two polarizations of 
$\abs{\Delta \phi} = \lr{2\pi/\lambda } \abs{\Delta n} d =4.2$ deg.

\section{Absorption by NV centers}

NV centers are excited from the spin-triplet ground state $^3A_2$ to the spin-triplet excited state $^3E$ by 532 nm pump light. For small pump light intensities $I_P$ where the transition is not saturated, the excitation rate $\Gamma_P$ is 
\begin{equation}
\Gamma_P = \sigma_P \cdot \Phi_P = \sigma_P \cdot \frac{ I_P}{h c/\lambda_P},
\end{equation}
where $\sigma_P$ is the cross-section for the pump light on the $^3A_2 \leftrightarrow {^3E}$ transition, 
$\Phi_P$ is the flux of pump photons, $h$ is Planck's constant, and $\lambda_P$ is the pump wavelength. 
The condition for saturating the $^3A_2 \leftrightarrow {^3E}$ transition is $\Gamma_P^{\rm{sat,triplet}} \approx A_{21}$, where $A_{21}$ is the inverse lifetime of the ${^3E}$ excited state. From this condition we calculate the saturation intensity to be 
\begin{equation}
I_P^{\rm{sat,triplet}} \approx  \frac{A_{21}}{\sigma_P} \cdot \frac{hc}{\lambda_P} \sim 10~\mathrm{GW/m}^2 ,
\end{equation}
where we have used the values $A_{21} =1/\lr{12~\mathrm{ns}}$ \cite{Batalov2008}, $\sigma_P=3\cdot10^{-21}~\mathrm{m}^2$ \cite{Wee2007}, and $\lambda_P=532$~nm.\\

An NV center can decay from the spin-triplet excited state $^3E$ to the upper spin-singlet state $^1A_1$ from which it   decays in $<1$~ns \cite{Acosta10PRB} to the lower metastable spin-singlet state~$^1E$.
The NV center will therefore mainly populate the lower  spin-singlet state under strong optical pumping. 
The saturation pump power for populating the singlet state can be calculated using a rate-equation model for the spin-state dynamics of the NV center \cite{Robledo11}.
An order of magnitude estimate can be obtained from the condition $\Gamma_P^{\mathrm{sat,singlet}} \approx \Gamma_s$, where $\Gamma_s=1/\lr{200~\rm{ns}}$ is the inverse lifetime of the metastable singlet state \cite{Acosta10PRB}. 
The saturation intensity is then
\begin{equation}
I_P^{\rm{sat,singlet}} \approx  \frac{\Gamma_s}{\sigma_P} \cdot \frac{hc}{\lambda_P} = 
\frac{\Gamma_s}{A_{21}} \cdot I_P^{\rm{sat,triplet}}= 600~\mathrm{MW/m}^2 .
\end{equation}
The population in the metastable singlet state as a function of pump intensity can be estimated as 
\begin{equation}
p_s \approx \frac{I_P}{I_P+I_P^{\mathrm{sat,singlet}}}.
\end{equation}
We now consider a diamond with a density of NV centers $\nNV$ and thickness $d$. If we assume that all NV centers experience the same pump intensity $I_P$, the amount of absorption is
\begin{equation}
A = \sigmaIR \cdot  \nNV \cdot d \cdot p_s
   =  \sigmaIR \cdot  \nNV \cdot d \cdot  \frac{I_P}{I_P+I_P^{\mathrm{sat,singlet}}},
\end{equation}
where $\sigmaIR$ is the cross-section for the $^1E \leftrightarrow {^1A_1}$ transition.
In our experiment the pump intensity is not homogeneous inside the diamond since the pump light has a Gaussian transverse intensity profile and because the pump light is attenuated while passing through the diamond due to absorption by NV centers. 
However, for simplicity, we assume a homogeneous pump intensity in our model used for fitting the experimental data
shown Fig.~2(b)
for the cavity transmission as a function of pump power.

\begin{figure}
\centering
\includegraphics[width=0.4\textwidth]{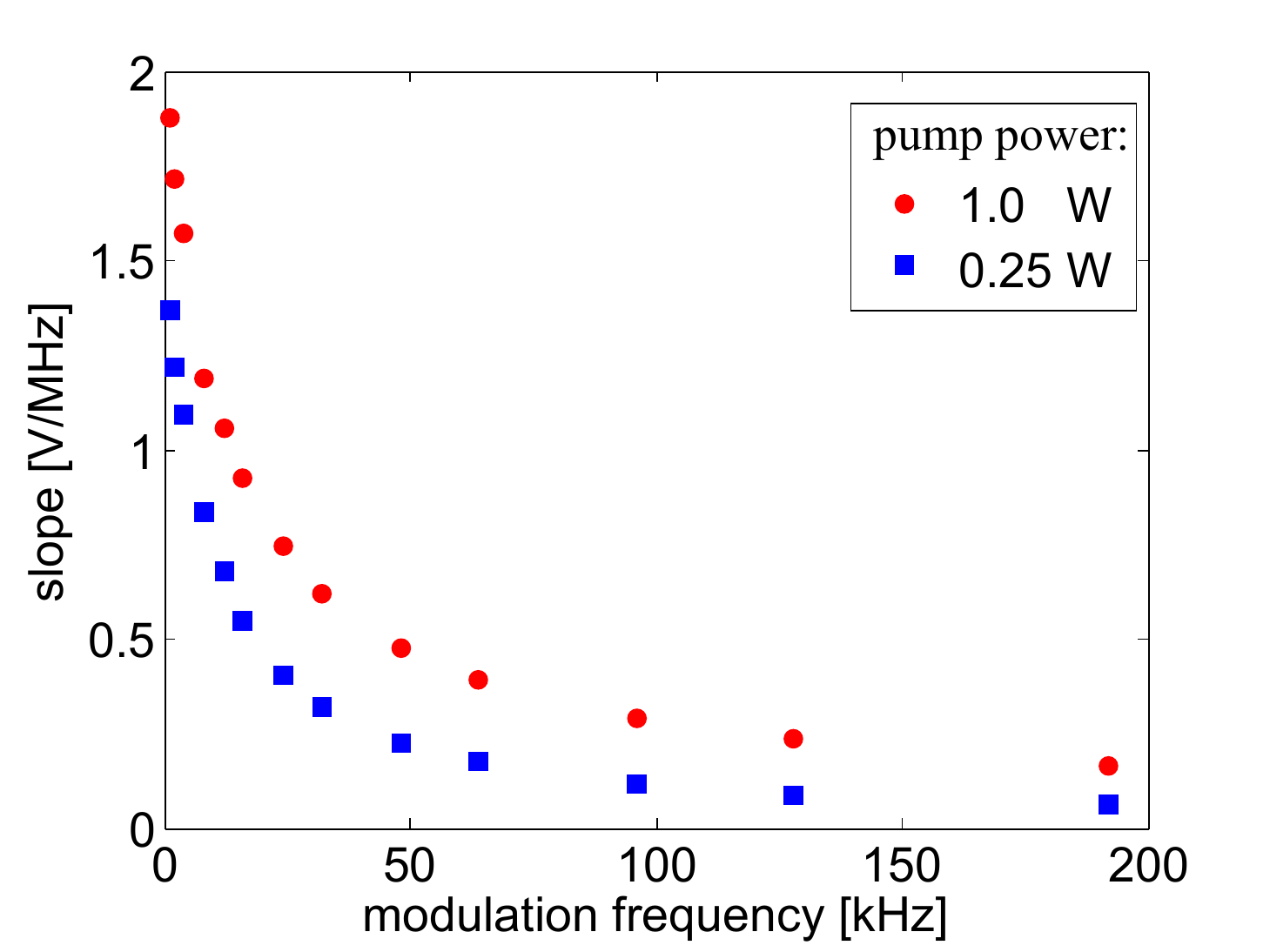}
\caption{The slope $\alpha$ as a function of the modulation frequency for two pump-light powers 
(1.0~W and 0.25~W measured before the cavity).
}
\label{fig:bandwidth}
\end{figure}

\section{Magnetometer bandwidth}
Using the modulation technique described in the main text, we obtain a magnetometer signal 
$ \SLI \approx  \alpha \lr{f_c-\fres}$ which is linear in $f_c-\fres$ close to resonance.
Figure~\ref{fig:bandwidth} shows the slope $\alpha$ for different modulation frequencies.
%Figure~\ref{fig:bandwidth} shows the linear part of the magnetometer's response, the slope $\alpha$, for different modulation frequencies.
The bandwidth of the magnetometer, defined as the modulation frequency where $\alpha$ has decreased by a factor of two, is 13.5~kHz and 10.6~kHz for the pump powers $1.0$~W and $0.25$~W, respectively. Those powers correspond to peak intensitites 80~MW/$\mathrm{m}^2$ and 20~MW/$\mathrm{m}^2$, respectively (with a 90~$\upmu$m beam waist).
Higher bandwidths can be achieved by increasing the pump intensity. In Ref.~\cite{Shin2012}, a bandwidth of a few MHz was achieved with a pump intensity of 20~GW/$\mathrm{m}^2$. 
The maximum bandwidth of IR-absorption based magnetometry in diamond is limited by the metastable singlet state lifetime to a few MHz \cite{Acosta10APL}.

%\bibliography{BIBInfraredAbsorption} 
%\bibliographystyle{unsrt} 

%%%%%%%%%%%%%%%%% bibliography %%%%%%%%%%%%%%%%%%%

%%%%%%%%%%%%%%%%%%%%%%%%%%%%%%%%%%%%%%%%%%%%%%%%%%